\documentclass[prb, aps, twocolumn, amsmath, amssymb, superscriptaddress]{revtex4-1}
\usepackage{graphicx,verbatim}
\usepackage{braket}
\usepackage{color}

\usepackage{soul}
\usepackage[normalem]{ulem}
\usepackage[x11names]{xcolor}

\begin{document}

\title{Revealing strong correlations in higher order transport statistics: a noncrossing approximation approach}

\author{A.\ Erpenbeck}
	\affiliation{
	The Raymond and Beverley Sackler Center for Computational Molecular and Materials Science, Tel Aviv University, Tel Aviv 6997801, Israel
	}
	\affiliation{
	School of Chemistry, Tel Aviv University, Tel Aviv 6997801, Israel
	}

\author{E.\ Gull}
	\affiliation{
	Department of Physics, University of Michigan, Ann Arbor, Michigan 48109, USA
	}

\author{G.\ Cohen}
	\affiliation{
	The Raymond and Beverley Sackler Center for Computational Molecular and Materials Science, Tel Aviv University, Tel Aviv 6997801, Israel
	}
	\affiliation{
	School of Chemistry, Tel Aviv University, Tel Aviv 6997801, Israel
	}

\date{\today}

\begin{abstract}
	We present a method for calculating the full counting statistics of a nonequilibrium quantum system based on the propagator noncrossing approximation (NCA).
	This numerically inexpensive method can provide higher order cumulants for extended parameter regimes, rendering it attractive for a wide variety of purposes.
	We compare NCA results to Born--Markov quantum master equations (QME) results to show that they can access different physics, and to numerically exact inchworm quantum Monte-Carlo data to assess their validity. 
	As a demonstration of its power, the NCA method is employed to study the impact of correlations on higher order cumulants in the nonequilibrium Anderson impurity model.
	The four lowest order cumulants are examined, allowing us to establish that correlation effects have a profound influence on the underlying transport distributions.
	Higher order cumulants are therefore demonstrated to be a proxy for the presence of Kondo correlations in a way that cannot be captured by simple QME methods.
	
\end{abstract}

\maketitle

\section{Introduction}\label{sec:introduction}

\subsection{Background and overview}
Electron transport through mesoscopic and nanoscale junctions is a complex phenomenon where nonequilibrium statistical mechanics is entwined with quantum many-body effects.\cite{datta_electronic_1997, bruus_many_2004, stefanucci_nonequilibrium_2013, cohen_greens_2020}
Systems are driven out of equilibrium by, e.g., an external bias voltage or a temperature gradient, and their response is measured.
Perhaps the simplest response observable in many experimental setups is the electronic current.
Increasingly, however, is has become both possible and desirable to access so-called higher order transport characteristics.
This includes the current's fluctuations and its higher moments\cite{reulet_environmental_2003, bomze_measurement_2005, Gustavsson2006} 
as well as the statistics of individual electron transfer events.\cite{kung_irreversibility_2012}
Interestingly, ultracold atom experiments can simulate electronic transport,\cite{brantut_conduction_2012} and allow for directly extracting statistical distributions of populations in different parts of the system,\cite{mazurenko_cold-atom_2017} marking another path towards detailed characterization of transport.

Theoretically, all such information can be obtained from the full counting statistics (FCS) approach pioneered Levitov and Lesovik,\cite{Levitov1993, Levitov1996} where all moments and cumulants of transport events are efficiently represented by a single generating functional.
Since its inception, this idea has attracted a great deal of attention.\cite{Bagrets2003, Belzig2005, Flindt2005_2, koch_full_2005, Esposito_Fluctuation_2007, Esposito_Entropy_2007, Esposito2009, Xue2011, Nicolin_Non_2011, Kambly2013, Kaasbjerg2015, Ridley2018, Ridley2019, Schinabeck2020, Kurzmann_Optical_2019}

Many experimental studies concentrate on the current noise and the current-to-noise ratio, also known as the Fano factor.
In both classical and quantum systems, these quantities already contain information not present in the mean current:\cite{Landauer1998, Blanter2000} for example, they enable probing of effective quasiparticle charges.\cite{deJong1994, DePicciotto1997, Lefloch2003}
Moreover, noise measurements have allowed researchers to, e.g., identify electron bunching and anti-bunching during transport;\cite{Blanter2000, Safonov2003, Djuric2005, Gustavsson2006, Tworzydlo2006, Kiesslich2007, Emary2007} reconstruct waiting- and dwell-time distributions;\cite{Beenakker2003, Tang2014, Rudge2016, Ptaszynski2017, Kosov2018, Ridley2018, Stegmann_Real_2020} and determine the number and transmission probabilities of active levels contributing to transport\cite{vandenBrom1999, Cron2001, Djukic2006, Kiguchi2008, Tal2008, Wheeler2010, Schneider2010, Vardimon2013}.
Other studies reported the measurement of higher order cumulants that further elucidate the mechanisms underlying electronic transport.\cite{Flindt2009, Fricke2010, Ubbelohde2012}

Given sufficient cumulants, it is in principle possible to reconstruct the full FCS.
Much of the motivation for this comes from insights regarding noninteracting systems, where the exact FCS is given by the Levitov--Lesovik formula.\cite{Levitov1996, Nazarov2009}
There, the ability to measure the FCS could provide indirect access to theoretically intuitive but experimentally unattainable properties like channel coherence\cite{Brandes2008} and entanglement entropy.\cite{Klich_Levitov_2009}
This scheme holds also true for interacting systems, where the FCS provides insight onto many-body quantum effects.
For example, even though the role of electronic correlations is not yet well understood, it is known that correlation-driven physics like the Kondo effect modify the current noise\cite{Hewson1997kondo, Meir2002, Delattre2009} and its higher order cumulants.\cite{Stegmann_Detection_2015, Ridley2019}
Still, the theoretical prediction of the FCS for interacting systems is generally non-straightforward and 
a variety of theoretical approaches has been applied.
Among the approximate approaches used are quantum master equations (QME),\cite{Bagrets2003, Flindt2005_2, Flindt2008, Brandes2008, Flindt2010, Albert2011, simine_vibrational_2012, Schinabeck2014, Kaasbjerg2015, Benito2016, Stegmann2017, Kosov2017, Lead_Geometry} and Green's function based approaches.\cite{Tang2014, Galperin2006, Avriller2009, Schmidt2009, Haupt2010, Novotny2011, Utsumi2013, agarwalla_full_2015, Miwa2017, Stadler2018, cohen_greens_2020}
Numerically exact approaches to FCS include the Inchworm quantum Monte Carlo (iQMC) method,\cite{cohen_taming_2015,Ridley2018, Ridley2019} the hierarchical equations of motions technique (HEOM),\cite{Cerrillo2016, Schinabeck2020} the density matrix renormalization group approach\cite{Carr2011, Schmitteckert2014, Carr2015} and the iterative path integral method.\cite{weiss_iterative_2008,segal_numerically_2010,simine_vibrational_2012,agarwalla_full_2015,kilgour_path-integral_2019}
A variety of ongoing research programs are aimed at extending exact approaches to new experimentally relevant regimes, and at developing new exact and approximate methodologies.

\subsection{Noncrossing approximations}
At the present time, methods able to address Kondo physics remain computationally expensive.
Here, we propose a simple and inexpensive approximate scheme for evaluating FCS that is based on one variation of the noncrossing approximation (NCA).
The NCA and its extensions\cite{Bickers1987,haule_anderson_2001,Eckstein2010} have long been a successful qualitative approach to several aspects of nonequilibrium Kondo physics in quantum transport.\cite{meir_low-temperature_1993, Wingreen_Anderson_1994,hettler_nonequilibrium_1998,nordlander_how_1999,plihal_transient_2005,hartle_decoherence_2013,chen_anderson-holstein_2016,roura-bas_nonequilibrium_2013,Peronaci_Resonant_2018,krivenko_dynamics_2019,atanasova_correlated_2020,Erpenbeck_Resolving_2020}
The approximation has multiple, inequivalent formulations, most of which are unsuitable to the evaluation of FCS due to the introduction of an auxiliary pseudoparticle space.\cite{cohen_greens_2020}
The formulation used here is a lowest order precursor of the hybridization-expansion-based iQMC method,\cite{cohen_taming_2015,antipov_currents_2017} and the starting point of bold-line schemes that preceded it.\cite{gull_bold-line_2010,gull_numerically_2011,cohen_numerically_2013,cohen_greens_2014,cohen_greens_2014-1}
It can easily be used to obtain high order cumulants or the complete FCS generating functional.
To highlight the advantages of the NCA, we contrast it with the widely applied QME scheme, which completely neglects Kondo physics.\cite{Miwa2017}
We then establish that our NCA provides better results than the QME scheme by comparing with numerically exact data obtained from iQMC.
Finally, based on the NCA, we provide a preliminary overview of the signature of nonequilibrium correlation effects in higher order cumulants.

\subsection{Quantum master equations}\label{sec:QME}
One of the two methods to which we will provide direct comparisons is the QME approach.
Similarly to the NCA to be presented below in Sec.~\ref{sec:methodology}, the QME approximation is based on a second order expansion in the dot--lead coupling.
In contrast to the NCA, the QME does not employ a Dyson-like diagrammatic resummation scheme.
Rather, it uses a Liouville-space resummation based on the Nakajima--Zwanzig equation.\cite{Nakajima1958, Zwanzig1960, Fick1990quantum}
This results in an analytically solvable and intuitive equation of motion for the reduced density matrix, which is the method of choice in many contexts.\cite{Erpenbeck_W_Term, Harbola_Quantum_2006, Peskin_2010, Haertle2011, simine_vibrational_2012, Schinabeck2014, Kaasbjerg2015, Kosov2017, Purkayastha_Quantum_2017, Lead_Geometry}

QME methods have been widely employed in the evaluation of FCS.\cite{Bagrets2003,Lead_Geometry, Flindt2005_2, Kaasbjerg2015, rudge_fluctuating_2019}
Their numerically exact generalization, the HEOM technique,\cite{Tanimura1989, Tanimura2006, Jin2008, Zheng2012, hartle_decoherence_2013, Schinabeck2016, Erpenbeck_RSHQME, Erpenbeck_Hierarchical_2019, Tanimura2020, Erpenbeck_Current_2020}
has recently been generalized to FCS in the context of vibrationally coupled electronic transport.\cite{Schinabeck2020}

\subsection{Inchworm quantum Monte Carlo method}\label{sec:iQMC}
The inchworm quantum Monte Carlo (iQMC) method
is a numerically exact framework able to evaluate transport properties in correlated nonequilibrium impurity models.\cite{cohen_taming_2015, Chen2017, Chen2017_2, antipov_currents_2017, Dong2017,  Boag2018, cai_inchworm_nodate, cai_numerical_2020, eidelstein_multiorbital_2020}
It has recently been used to evaluate the FCS of both particle and energy transport in the presence of electron--electron interactions.\cite{Ridley2018, Lead_Geometry, Ridley2019}

In the present context, the iQMC framework can be considered a numerically exact generalization of the NCA method.
This does not mean that the NCA is immediately obsolete, just as the availability of HEOM methods has not obviated QME approximations.
This is natural because iQMC results are substantially more expensive to obtain than NCA results, especially at steady state.
Here, we employ the iQMC method to validate our NCA results and illustrate their usefulness.

\subsection{Outline of this work}
We will proceed as follows: 
In Sec.~\ref{sec:model}, we introduce the model system investigated in this work.
The FCS formalism is outlined in Sec.~\ref{sec:FCS}.
The theoretical NCA framework employed in this work is described in Sec.~\ref{sec:methodology}.
In Sec.~\ref{sec:results}, we present our results: comparisons between NCA and QME results are given in \ref{sec:NCA_VS_QME}, physical implications are discussed in \ref{sec:Fano}, and validation with respect to iQMC is presented in \ref{sec:NCA_VS_iQMC}.
Finally, in Sec.~\ref{sec:summary}, we conclude and summarize our findings.

\section{Model}\label{sec:model}
We consider the nonequilibrium Anderson impurity model, a minimal description for a finite, interacting quantum dot coupled to two infinite noninteracting leads.
The Hamiltonian is
\begin{eqnarray}
H 	&=& 	H_D + H_B + H_{DB} , \label{eq:H_full}
\end{eqnarray}
where $H_D$ is in the dot subspace, $H_B$ is in the bath subspace comprising the left and right lead, and $H_{DB}$ encodes the coupling between the dot and the leads.	

The dot Hamiltonian is given by
\begin{eqnarray}
H_D	&=& 	\sum_{\sigma=\uparrow,\downarrow} \epsilon_0 d_\sigma^\dagger d_\sigma + U d_\uparrow^\dagger d_\uparrow d_\downarrow^\dagger d_\downarrow. \label{eq:H_D}
\end{eqnarray}
Here, the $d_\sigma^{\left( \dagger\right)}$ denote creation/annihilation operators for an electron of spin $\sigma$ on the dot, $\epsilon_0$ is the single particle occupation energy, and $U$ determines the strength of the Coulomb interaction.
Experimentally, the single particle occupation energy can be tuned by an external gate voltage $\Phi_{\text{gate}}$.
We model the influence of such a gate voltage by setting $\epsilon_0 = \Phi_{\text{gate}} - \frac{U}{2}$.

The leads are assumed to be a noninteracting continuum,
\begin{eqnarray}
H_B	&=&	\sum_{\sigma\in\lbrace\uparrow,\downarrow\rbrace}\sum_{\ell\in\lbrace L,R \rbrace}  \sum_{k\in\ell}   \epsilon_{k} a_{k\sigma}^\dagger a_{k\sigma} ,
\end{eqnarray}
where the $a_{k\sigma}^{(\dagger)}$ are creation/annihilation operators on a lead level with index $k$, spin $\sigma$ and energy $\epsilon_{k}$. The indices $L$ and $R$ denote the ``left'' and ``right'' lead, respectively.
Finally, the coupling between the dot and leads is assumed to take the linear form
\begin{eqnarray}
H_{DB}	&=&	\sum_{\sigma\in\lbrace\uparrow,\downarrow\rbrace} \sum_{\ell\in\lbrace L,R \rbrace}\sum_{k\in \ell}  \left( V_{k} a_{k\sigma}^\dagger d_\sigma + \text{h.c.} \right) , \label{eq:H_coupl}
\end{eqnarray}
with coupling parameters $V_{k}$ that can be parameterized in terms of a coupling strength function
\begin{eqnarray}
\Gamma_{\ell}(\epsilon)	&=&	\pi \sum_{k\in \ell} |V_{k}|^2 \delta(\epsilon-\epsilon_{k}).
\end{eqnarray}
We explicitly consider symmetric coupling to the two leads, each of which is taken to be a flat band with a soft cutoff:
\begin{equation}
\Gamma_{L}(\epsilon)=\Gamma_{R}(\epsilon)=\frac{\Gamma/2}{(1+e^{\nu(\epsilon-\epsilon_c)})(1+e^{-\nu(\epsilon+\epsilon_c)})}.
\end{equation}
The overall strength of the dot--lead coupling is set by the constant $\Gamma$, which is used as our unit of energy.
The coupling strength defines the hybridization functions, 
\begin{eqnarray}
	\Delta_{\ell}^<(t)	&=&	\frac{1}{\pi}\int d\epsilon\ e^{+i\epsilon t}\ \Gamma_{\ell}(\epsilon) f_\ell(\epsilon) , \\
	\Delta_{\ell}^>(t)	&=&	\frac{1}{\pi}\int d\epsilon\ e^{-i\epsilon t}\ \Gamma_{\ell}(\epsilon) (1-f_\ell(\epsilon)). \label{eq:def_Delta_>}
\end{eqnarray}
Here $f_\ell(\epsilon) \equiv \frac{1}{1+e^{\beta (\epsilon-\mu_\ell)}}$, where $\mu_{{L/R}}= \pm V/2$ are chemical potentials set by a symmetrically applied bias voltage $V$, and $\beta$ is the inverse temperature in the leads.
Moreover, we set $\nu=1/\Gamma$ and $\epsilon_c=50\Gamma$ -- much larger than all other energy scales in the problem -- such that we are effectively working in the wide band limit.
With our choice of parameters, particle-hole symmetry is obeyed for $\Phi_{\text{gate}}=0$.

Throughout this work, the on-site Coulomb repulsion is set to $U=8\Gamma$.
This determines a Kondo temperature of $T_K \approx 0.8\Gamma$ for the particle--hole symmetric case,\cite{Hewson1997kondo} which we use as a reference for the emergence of the Kondo phenomenon.
Generally, the Kondo temperature depends on the gate voltage\cite{Hewson1997kondo}  and we will comment on this at appropriate points below.
We will consider three representative lead temperatures: $T=0.25\Gamma<T_K$, $T=0.5\Gamma \lesssim T_K$ and $1.0\Gamma \gtrsim T_K$, whereby $T_K$ is the estimate for the Kondo temperature for the particle-hole symmetric scenario.
This means that we are exploring the edge of the Kondo regime rather than the deep Kondo regime where scaling behavior can be extracted.
This choice is to some extent motivated by the limitations of the methods used in this work (cf.\ Sec.~\ref{sec:methodology}).

\section{FCS and counting fields}\label{sec:FCS}

Determining the FCS of an observable means evaluating the generating function of its underlying probability distribution, from which cumulants and moments can be extracted.
We provide a brief overview of this approach and the main concepts here, and recommend Refs.~\onlinecite{Esposito2009, Utsumi2019} for more details.

Consider an experiment where at time zero the system is prepared in a known initial density matrix where, e.g., the number of electrons in the left lead $L$ is known.
The system is allowed to evolve freely until time $t$, when the total number of electrons in lead $L$ is measured.
Let $P_L(t,n)$ be the probability that $n$ electrons are found in this measurement.
The generating function is then defined as
\begin{equation}
Z_L(t, \lambda)	\equiv	\sum_n P_L(t,n) e^{i\lambda n} \equiv \text{Tr}_{D+B} \left\lbrace \rho_\lambda(t) \right\rbrace , \label{eq:general_def_Z}
\end{equation}
where $\lambda$ is known as the counting field.
This defines $\rho_\lambda(t) \equiv e^{-iH_\lambda t} \rho(0) e^{iH_{-\lambda}t}$, a counting-field-modified (or, for brevity, simply ``modified'') density matrix; which in turn defines $H_\lambda \equiv e^{i\lambda/2 N_L} H e^{-i\lambda/2 N_L}$, a modified Hamiltonian.
$N_L = \sum_{\sigma\in\lbrace\uparrow,\downarrow\rbrace}\sum_{k\in L} a_{k\sigma}^\dagger a_{k\sigma}$ is the particle number operator in the left lead $L$.
Modifying the Hamiltonian by the counting field corresponds to transforming the dot--bath coupling strength of the lead under consideration according to\cite{tang_full-counting_2014}
\begin{equation}
	V_{k}(t_\pm)	\rightarrow	V_{k} e^{\pm i\lambda/2}, \label{eq:dressing_HI}
\end{equation}
Where $t_\pm$ is a time variable on either the backward ($+$) or forward ($-$) branch of the Keldysh contour.
This idea can be generalized to other observables and counting fields.\cite{Esposito2009}

Normally, the generating function $Z_L(t, \lambda)$ itself cannot be directly accessed in experiments.
However, experiments can measure its moments and cumulants, or sometimes the probabilities $P_L(t, n)$.
In particular, the cumulants $C_L^\alpha(t)$ of the generating function are given by its logarithmic derivatives:
\begin{eqnarray}
	C_L^\alpha(t)	&=&	(-i)^{\alpha}\frac{\partial^\alpha}{\partial\lambda^\alpha} \ln\left(Z_L(t, \lambda)\right) \Big|_{\lambda=0}. \label{eq:def_cumulant}
\end{eqnarray}

The first few cumulants have simple physical interpretations.
The time derivative of the first cumulant corresponds to the electronic current $I_L(t)$ exiting lead $L$:
\begin{eqnarray}
	C_L^1(t)				&=&	\braket{N_L(t)}, \\
	\frac{\partial}{\partial t} C_L^1(t)	&=&	I_L(t).
\end{eqnarray}
The second cumulant is related to the variance of the population in the lead,
\begin{eqnarray}
	C_L^2(t)	 			&=&	\braket{N_L^2}(t) - \braket{N_L}^2(t).
\end{eqnarray}
At steady state, its time derivative is the noise $S_L$:
\begin{equation}
\lim_{t\rightarrow\infty}\frac{\partial}{\partial t} C_L^2(t) = S_L.
\end{equation} 
Higher order population cumulants and the full probability distributions $P_L(t,n)$ can also be obtained from the generating functional.
These have a more complicated relationship with the statistics of the current, but are arguably more straightforward than the latter to describe theoretically.
For the scope of the this work, we will also consider the steady state time derivatives of the third and fourth cumulants, $\lim_{t\rightarrow\infty}\frac{\partial}{\partial t} C_L^3(t) \equiv S_{L2}$ and $\lim_{t\rightarrow\infty}\frac{\partial}{\partial t} C_L^4(t) \equiv S_{L3}$. 
These quantities express the skewness and the bifurcation of the underlying probability distribution, respectively, and are of interest in a variety of contexts.\cite{Belzig2005, Xue2011}
Composite observables like the Fano factor $F_L=S_L / I_L$ are often easier to obtain experimentally than the cumulants themselves, because they do not vary with the overall conductivity of the junction.

The standard Fano factor can be a problematic quantity for studying Kondo physics, because the low energy features are obscured by the zero bias Nyquist--Johnson singularity.\cite{Blanter2000, CuevasScheer}
This stems from the different symmetry of $I$ and $S$ with respect to the bias voltage. 
Deep in the universal Kondo regime and at very low voltages, this can be rectified by defining a ``backscattering'' current that must be separated from the unitary linear-response current.\cite{sela_fractional_2006,ferrier_universality_2016}
Below, we discuss an alternative and more widely applicable approach: to define a set of generalized Fano factors in terms of higher order cumulants, while taking symmetry into account.

\section{Methodology}\label{sec:methodology}
\subsection{Noncrossing approximation}\label{sec:NCA_Method}

NCA refers to a wider class of inequivalent methods that are perturbative in the dot-bath coupling. The name is motivated by the fact these methods only consider contributions to the perturbative series, which have a diagrammatic representation in which the hybridization lines do not cross. The first installment of an NCA method roots back to Grewe and Kuramoto\cite{Grewe_Diagrammatic_1981, Kuramoto_Self_1983} and was employed and extended by various authors to account for finite electron-electron interaction strengths\cite{Pruschke_Anderson_1989, Keiter_The_1990} and nonequilibrium conditions.\cite{Wingreen_Anderson_1994}
The corresponding formulation of the NCA uses a pseudoparticle representation in order to make quantum field theoretical methods such as Wick's theorem applicable. This, however, enlarges the underlying Hilbert space into unphysical regions and relies on a representation where the number of electrons on the dot is not well defined at any given time. Consequently, in this formulation of the NCA, the evaluation of FCS it is not straightforward.
Still, this NCA scheme is well suited to capture physics at temperatures that are not far below the Kondo temperature and works well in the large $U$ limit and for small bias voltages.
The minimal NCA does not correctly capture Kondo physics in the scaling regime at quantitative accuracy, but this can be amended to a large degree with the aid of vertex corrections.\cite{Anders_Beyond_1995, Anders_Perturbational_1994, Grewe_Conserving_2008, Eckstein2010}
These more advanced, but expensive, extensions of the NCA method have been successfully employed in recovering the temperature scaling behavior characterizing Kondo phenomena in agreement with numerical renormalization group calculations.\cite{Gerace_Low_2002, Kroha_Conserving_2005}

In the present work, we employ a different NCA scheme, which is based on the perturbative expansion of the restricted propagator (cf.\ Eq.~(\ref{eq:def_res_propagator})) in terms of the dot-bath coupling\cite{cohen_numerically_2013, cohen_greens_2014, chen_anderson-holstein_2016} and which represents a precursor of the QMC methods based on the hybridization expansion.\cite{cohen_taming_2015,antipov_currents_2017, gull_bold-line_2010, gull_numerically_2011, cohen_numerically_2013, cohen_greens_2014-1}
This NCA formulation employs the occupation number basis of the interacting dot, such that the number of electrons on the dot is a well defined quantity at any given time. This allows for a straightforward calculation of the FCS using the transformation in Eq.\ (\ref{eq:dressing_HI}), while on the downside, tools like Wick's theorem are not applicable.
We will now describe the details of the propagator hybridization expansion for the FCS generating function $Z(t,\lambda) \equiv Z_L(t,\lambda)$ within the NCA.
The approximation is based on a second order expansion of the time evolution operator in the dot--lead coupling, which is treated self consistently within a Dyson resummation scheme.

Using Eq.~\eqref{eq:general_def_Z} in the context of Sec.~\ref{sec:model} and assuming an initial condition factorized between the dot and bath spaces, $\rho(t=0,\lambda)=\rho_B\otimes\rho_D$, we obtain
\begin{equation}
	Z(t, \lambda)	=	\text{Tr}(\varrho(t, \lambda)) = 
	\sum_{\alpha\beta} \braket{\beta|\rho_D|\beta} K_{\alpha}^{\beta}(t,t,\lambda). \label{eq:Z_NCA}
\end{equation}
Here, $\alpha$ and $\beta$ are electron number states in the interacting dot, and the vertex function $K_{\alpha}^{\beta}(t,t', \lambda)$ takes the form
\begin{eqnarray}
	K_{\alpha}^{\beta}(t,t', \lambda)	&=&	\text{Tr}_B \left\lbrace \rho_B
								\bra{\alpha} U_{-\lambda}^\dagger(t) \ket{\beta}\bra{\beta} U_\lambda(t') \ket{\alpha}
								\right\rbrace . \nonumber \\ \label{eq:def:K_chi} 
\end{eqnarray}	
$\text{Tr}_B$ denotes tracing over the bath degrees of freedom.
We have also made use of a modified time evolution operator, $U_{\pm\lambda}(t) \equiv \mathrm{T}\exp(-i\int_0^t H_{\pm\lambda}(\tau) d\tau)$, where $\mathrm{T}$ is the time ordering operator.
The vertex function $K_{\alpha}^{\beta}(t,t', \lambda)$ is the central object within the specific NCA method used in this work.
In other contexts, without FCS, only the $\lambda=0$ form appears.
This can be used to construct approximate expressions for the expectation values of a variety of observables.\cite{Eckstein2010,antipov_currents_2017}

To derive the NCA, one starts with the perturbative expansion of Eq.~\eqref{eq:def:K_chi} in the dot--lead coupling $H_{DB}$,
\begin{equation}
\begin{aligned}
K_{\alpha}^{\beta}(t,t',\lambda) &=
 \sum_{n,m=0}^\infty \hspace*{-0.2cm} (i)^n (-i)^m 
 \int_0^t \hspace*{-0.2cm} d\tau_1 \dots  \int_0^{\tau_{n-1}} \hspace*{-0.6cm} d\tau_n 
 \int_0^{t'} \hspace*{-0.25cm} d\tau_1' \dots  \int_0^{\tau_{n-1}'} \hspace*{-0.6cm} d\tau_n'  \\ & \times
					\text{Tr}_B \Big\lbrace \rho_B
						\bra{\alpha} h_{-\lambda}(\tau_1) \dots h_{-\lambda}(\tau_n) e^{iH_0t} \ket{\beta}  \\&
						\hspace*{1cm}\times \bra{\beta} e^{-iH_0t'} h_{\lambda}(\tau_1') \dots h_{\lambda}(\tau_n') \ket{\alpha} 
					\Big\rbrace ,
\end{aligned}
\end{equation}
a diagrammatic representation of this expansion can for example be found in Refs.~\onlinecite{cohen_greens_2014, chen_anderson-holstein_2016}.
Here, $h(\tau) = e^{iH_0\tau} H_{DB} e^{-iH_0\tau}$ and $H_0 = H_D+H_B$.
The NCA is based on the lowest nonvanishing correction, which is then iterated until self consistency.
The approximation is obtained by expressing the vertex function in terms of this correction, resulting in the Dyson equation
\begin{equation}
\begin{aligned}
K_{\alpha}^{\beta}(t,t',\lambda)
&=k_{\alpha}^{\beta}\left(t,t^{\prime}\right)+\sum_{\alpha^{\prime}\beta^{\prime}}\int\limits _{0}^{t}\int\limits _{0}^{t^{\prime}}d\tau_{1}d\tau_{1}^{\prime} \\
&k_{\beta'}^{\beta}\left(t-\tau_{1},t'-\tau_{1}'\right)\xi_{\alpha'}^{\beta'}\left(\tau_{1}-\tau_{1}', \lambda\right)K_{\alpha}^{\alpha'}\left(\tau_{1},\tau_{1}',\lambda\right).
\end{aligned}
\label{eq:Dyson_K}
\end{equation}
This is defined in terms of the the cross-branch hybridization self-energy
\begin{eqnarray}
	\xi_{\alpha}^{\beta}(t, \lambda)
		&=&
			\sum_{\sigma\in \lbrace\uparrow, \downarrow\rbrace}
			\sum_{\ell\in\lbrace L,R \rbrace}
					\Big( 
					\Delta_{\ell}^<(t)
					e^{-i\lambda t}
					\braket{\alpha|d_\sigma|\beta} \braket{\beta|d_\sigma^\dagger|\alpha}
					\nonumber \\ &&
					+\Delta_{\ell}^>(t)
					e^{i\lambda t}
					\braket{\alpha|d_\sigma^\dagger|\beta} \braket{\beta|d_\sigma|\alpha}
					\Big),
\end{eqnarray}
and $k_{\alpha}^{\beta}\left(t,t^{\prime}\right)$, a term that is independent of the counting field and will be introduced momentarily.
The term NCA refers to the fact that there are no crossing hybridization lines in the diagrammatic representation of the terms included in this approach (see, e.g., Ref.~\onlinecite{cohen_greens_2014}).
Higher order expansions such as the one-crossing approximation employ different forms for the cross-branch self-energy.\cite{Pruschke_Anderson_1989,haule_anderson_2001,cohen_greens_2014}

We now return to the final quantity defined in Eq.~\eqref{eq:Dyson_K}, $k$.
This is a zeroth-order approximation for the vertex function that can be written in the form
\begin{equation}
k_{\alpha}^{\beta}(t,t') = \delta_{\alpha\beta} G_{\alpha}^*(t)G_{\beta}(t').
\end{equation}
Here,
\begin{equation}
	G_{\alpha}(t) = \braket{\alpha |\text{Tr}_B \left( \rho_B U_\lambda(t) \right) | \alpha}  \label{eq:def_res_propagator}
\end{equation}
is a single-branch propagator that is diagonal in the many-particle basis of the dot due to the structure of the Hamiltonian, Eq.~\eqref{eq:H_full}.
$G_{\alpha}(t)$ is also treated perturbatively in the dot--lead coupling,
\begin{equation}
\begin{aligned}
G_{\alpha}(t) &= g_{\alpha}(t) -
 \int_0^t \int_0^{\tau_{1}}  d\tau_1 d\tau_2 \\ &
					\hspace{1cm}
					\times
					\text{Tr}_B \Big\lbrace \rho_B
						\bra{\alpha} e^{-iH_0t} h_\lambda(\tau_1) h_\lambda(\tau_1) \ket{\alpha} 
					\Big\rbrace 
			+\dots, \label{eq:expansion_G}
\end{aligned}
\end{equation}
with $g_{\alpha}(t)=\braket{\alpha|e^{-i H_D t}|\alpha}$ being the propagator on the isolated dot.
We note that $G_\alpha(t)$ remains unmodified by the counting field, due to being restricted to one branch of the Keldysh contour such that all counting-field dependence on the right hand sinde of Eq.~(\ref{eq:expansion_G}) cancels out. 
Again, based on the lowest order of the expansion which is iterated until self consistency while neglecting diagrams that include hybridization lines,
$G$ obeys a set of equations similar to those obeyed by $K$, but on a single branch of the Keldysh contour:
\begin{equation}
\begin{aligned}G_{\alpha}(t) & = g_{\alpha}(t)\\
& -\int_{0}^{t}\int_{0}^{\tau_{1}}d\tau_{1}d\tau_{2}g_{\alpha}\left(t-\tau_{1}\right)\Sigma_{\alpha}\left(\tau_{1}-\tau_{2}\right)G_{\alpha}\left(\tau_{2}\right)
\end{aligned}
\label{eq:Dyson_G}
\end{equation}
The single-contour self-energy $\Sigma_{\alpha}(t)$ depends on the propagator $G_{\alpha}(t)$ and is given within the NCA by
\begin{eqnarray}
	\Sigma_{\alpha}(t)	&=&	
	\sum_{\sigma\in \lbrace\uparrow, \downarrow\rbrace}
	\sum_{\ell \in \lbrace L,R \rbrace}
	\sum_{\beta} \Big( 
				\Delta_{\ell}^<(t) \cdot \braket{\alpha | d_\sigma | \beta} \braket{\beta | d_\sigma^\dagger | \alpha} \nonumber \\&&
				+  \Delta_{\ell}^>(t) \cdot \braket{\alpha | d_\sigma^\dagger | \beta}  \braket{\beta | d_\sigma | \alpha} \Big) \cdot G_{\beta}(t) . \label{eq:def_Sigma_G}
\end{eqnarray}
Again, for a diagrammatic representation of this part of the expansion as well as the single-contour self-energies we refer to Refs.~\onlinecite{cohen_greens_2014, chen_anderson-holstein_2016}.

We conclude this section by commenting on the applicability of the NCA method, in particular to Kondo physics. 
Generally, the NCA is a method which is perturbative in the dot--lead coupling suggesting that its applicability is restricted to the strong interaction regime.
Still, its nonlinear nature makes its regime of validity hard to judge from simple analytical considerations.
This problem is exacerbated for nonequilibrium systems, systems with lower symmetry, and complex observables.
It has been argued that in equilibrium, the NCA provides accurate results for systems exhibiting strong interaction strengths $U$ as long as the temperature $T$ is not too low.\cite{Eckstein2010}
However, under nonequilibrium conditions, deviations from this rule have also been reported.\cite{cohen_numerically_2013}
Moreover, the method presented here is exact in the atomic limit, independent of the electron--electron interaction strength $U$.
It is formulated directly on the Keldysh contour, such that it is applicable to nonequilibrium conditions and not restricted to the linear response regime.
Nevertheless, it is known that NCA methods fail to provide accurate results in the small temperature limit. 
As such and as we noted in Sec.~\ref{sec:introduction}, it fails to provide accurate results for the scaling behavior or the Kondo temperature unless corrections are employed.
We therefore focus on higher-energy remnants of Kondo physics and nonequilibrium effects.
Even there, the treatment should be considered qualitative rather than quantitative.

	\subsection{Quantum master equations}\label{sec:QME2}

		Similar to the NCA approaches outlined in Sec.~\ref{sec:NCA_Method}, the QME method is based on a second order expansion in the dot--bath coupling.
		In contrast to NCA-based theories, the QME approach does not employ a Dyson scheme to incorporate a subset of diagrammatic contributions to the hybridization.
		Rather, it uses a Liouville-space resummation. 
		
		The QME is an equation of motion for the reduced density matrix of the dot, $\varrho(t) = \text{Tr}_B(\rho(t))$, where $\rho$ is the full density matrix of the dot and the bath and $\text{Tr}_B$ signifies a partial trace over the bath degrees of freedom.
		A formally exact equation of motion is provided by the Nakajima--Zwanzig equation.\cite{Nakajima1958, Zwanzig1960, Fick1990quantum}
		Expanding this to second order in the dot--bath coupling, in combination with the Markov approximation, results in the equation of motion:\cite{nitzan2013chemical}
		\begin{eqnarray}
			\frac{\partial}{\partial t}\varrho(t)
				&=&	-i [H_D, \varrho(t)] \\ && \nonumber
					\hspace*{-1.5cm}
					-
					\hspace*{-0.15cm}
					\int_0^\infty \hspace*{-0.35cm} d\tau \text{Tr}_L \hspace*{-0.1cm} \left( \hspace*{-0.05cm} \left[ H_{DB},  \hspace*{-0.1cm}\left[e^{-i(H_D+H_B)\tau} H_{DB} e^{i(H_D+H_L)\tau},   \hspace*{-0.05cm} \varrho(t) \rho_L  \hspace*{-0.05cm} \right]  \hspace*{-0.05cm} \right]  \hspace*{-0.05cm} \right)
					.
		\end{eqnarray}
		For the system under consideration, the populations and the coherences of the reduced density matrix decouple due to the form of Hamiltonians $H_D$ and $H_{DB}$ as given in Eqs.~\eqref{eq:H_D} and \eqref{eq:H_coupl}.
		As such, it is sufficient to consider the populations $p_\alpha(t) = \braket{\alpha|\varrho(t)|\alpha}$ of the reduced density matrix, whose dynamics obey the rate equations
		\begin{eqnarray}
			\frac{\partial}{\partial t} p_\alpha(t)	&=&	
				\sum_{\ell\in\lbrace L,R\rbrace \atop \beta\neq \alpha} |\vartheta_{\alpha\beta}| \times \label{eq:QME} \\ &&
					\Big(
					\Gamma_{\ell}(\vartheta_{\alpha\beta}(\epsilon_\alpha - \epsilon_\beta)) 
					f_\ell(\epsilon_\alpha - \epsilon_\beta)) 
					\times
					p_\beta(t) \nonumber \\ &&
					-  
					\Gamma_{\ell}(\vartheta_{\beta\alpha}(\epsilon_\beta - \epsilon_\alpha)) 
					f_\ell(\epsilon_\beta - \epsilon_\alpha)) 
					\times
					p_\alpha(t)\Big). \nonumber
		\end{eqnarray}
		The $\alpha$ and $\beta$ are, as before, states in the dot subspace;
		$n_\alpha$ and $n_\beta$ are the number of electrons residing on the dot in the state $\alpha$ and $\beta$, respectively; and $\vartheta_{\alpha\beta} = \pm 1$ if  $n_\alpha - n_\beta = \pm 1$, and zero otherwise.

		To obtains FCS, the populations are dressed by a counting field $p_\alpha(t) \rightarrow p_\alpha(t, \lambda)$.\cite{Bagrets2003}
		This corresponds to dressing the transition rates in Eq.~\eqref{eq:QME} according to
		\begin{eqnarray}
			\Gamma_{\ell}(\vartheta_{\alpha\beta}(\epsilon_\alpha - \epsilon_\beta))  
			&\rightarrow&
			\Gamma_{\ell}(\vartheta_{\alpha\beta}(\epsilon_\alpha - \epsilon_\beta)) e^{i\lambda\vartheta_{\alpha\beta}} .
		\end{eqnarray}
		The generating function is then calculated as the trace over the modified reduced density matrix, which is the sum over the modified populations:
		\begin{eqnarray}
			Z(t, \lambda)	&=&	\text{Tr}(\varrho(t, \lambda))
					= \sum_\alpha p_\alpha(t, \lambda).
		\end{eqnarray}

\section{Results}\label{sec:results}

Subsequently, we use the NCA methodology described above to study the four lowest order cumulants, $I_L$, $S_L$, $S_{L2}$, and $S_{L3}$, at steady state. As we are considering the steady state, we henceforth drop the lead index $L$.
Further, since these quantities diverge linearly in time, we plot their first time derivative.
We will investigate their dependence on bias voltage, gate voltage, and temperature.

\begin{figure*}[htb!]
  \centering
  \includegraphics{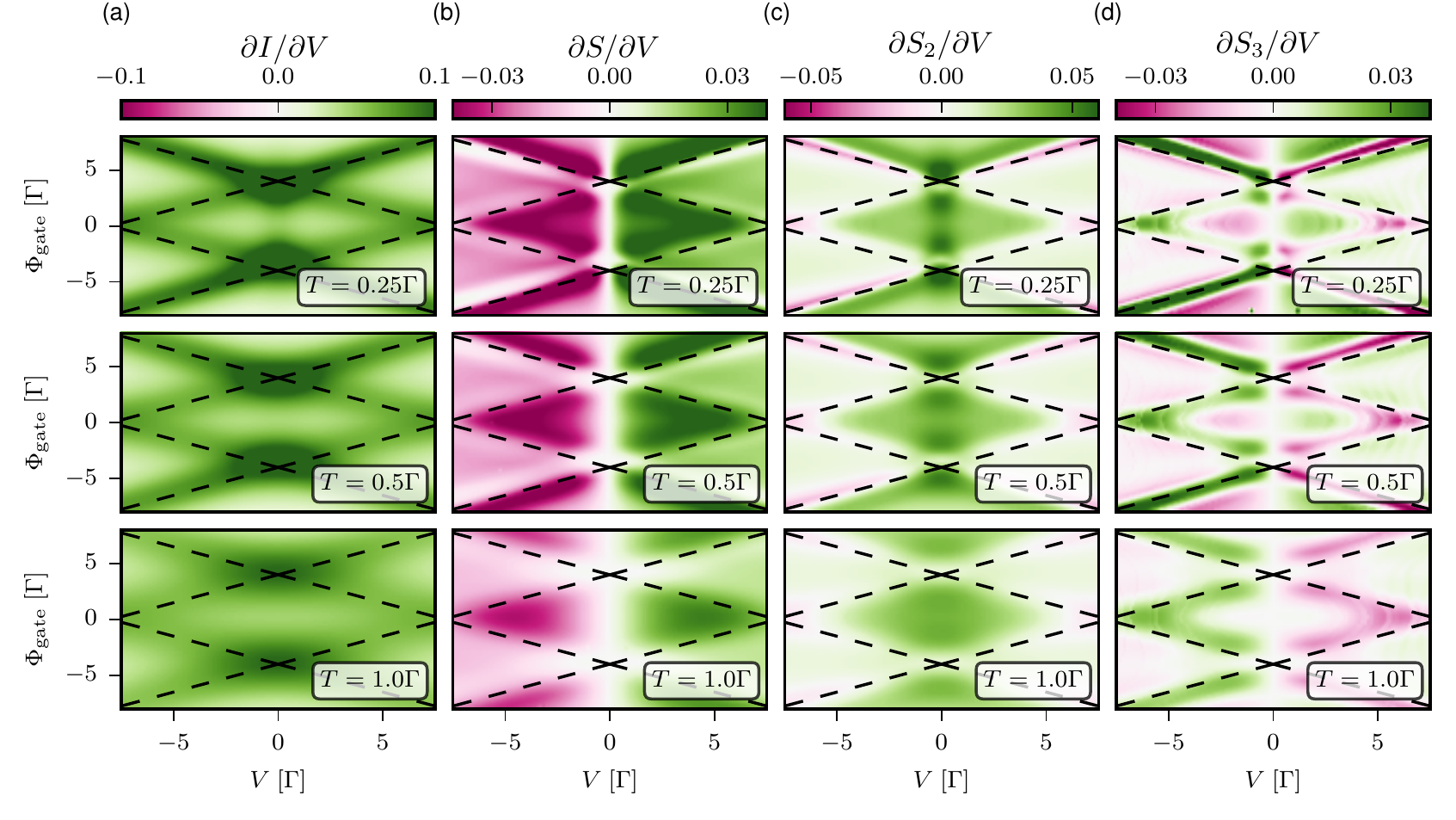}
  \caption{
		NCA results.
		The first derivative with respect to bias voltage is shown for the current (a), the noise (b), $S_2$ (c) and $S_3$ (d). From top to bottom, the temperature increases from $T=0.25\Gamma$ to $T=0.5\Gamma$ and finally $T=\Gamma$. 
		The black dashed lines, which serve as a guide for the eye, indicate the conditions $\epsilon_0 = \mu_{\text{L/R}}$ and $2\epsilon_0+U = \mu_{\text{L/R}}$ that separate resonant from nonresonant transport.
  \label{fig:NCA_mpas_dO}
	}
  \vspace*{0.305cm}
  \centering
  \includegraphics{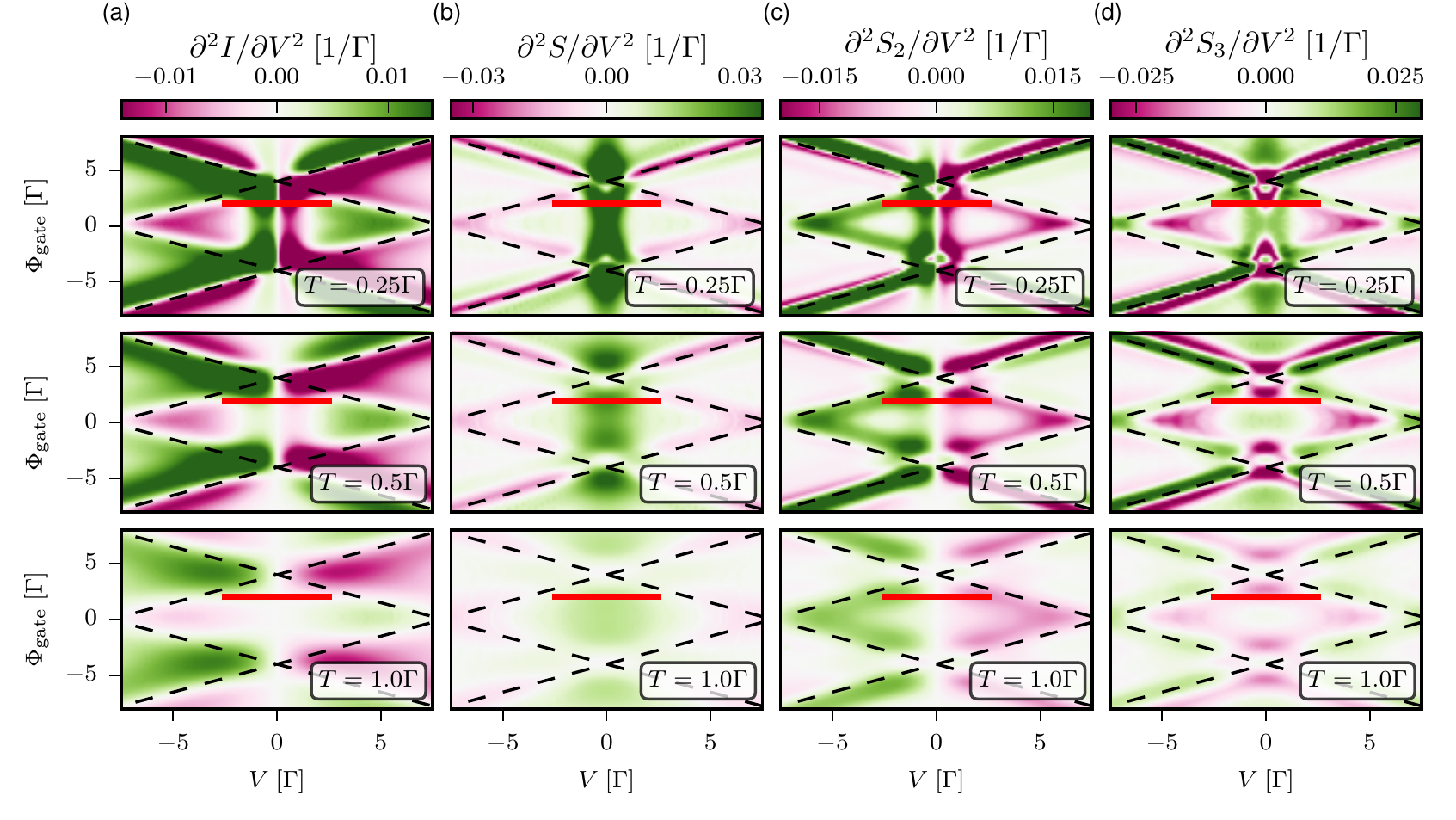}
  \caption{
		NCA results.
		The second derivative with respect to bias voltage is shown for the current (a), the noise (b), $S_2$ (c) and $S_3$ (d). From top to bottom, the temperature increases from $T=0.25\Gamma$ to $T=0.5\Gamma$ and finally $T=\Gamma$. 
		The black dashed lines, which serve as a guide for the eye, indicate the conditions $\epsilon_0 = \mu_{\text{L/R}}$ and $2\epsilon_0+U = \mu_{\text{L/R}}$ that separate resonant from nonresonant transport.
		Red solid lines indicate the parameters shown in Fig.~\ref{fig:cuts_ddO_comparison}.
		}
  \label{fig:NCA_mpas}
\end{figure*}
\subsection{Signature of correlations in observables associated to higher order cumulants}\label{sec:NCA_VS_QME}

\begin{figure*}[htb!]
  \centering
  \includegraphics{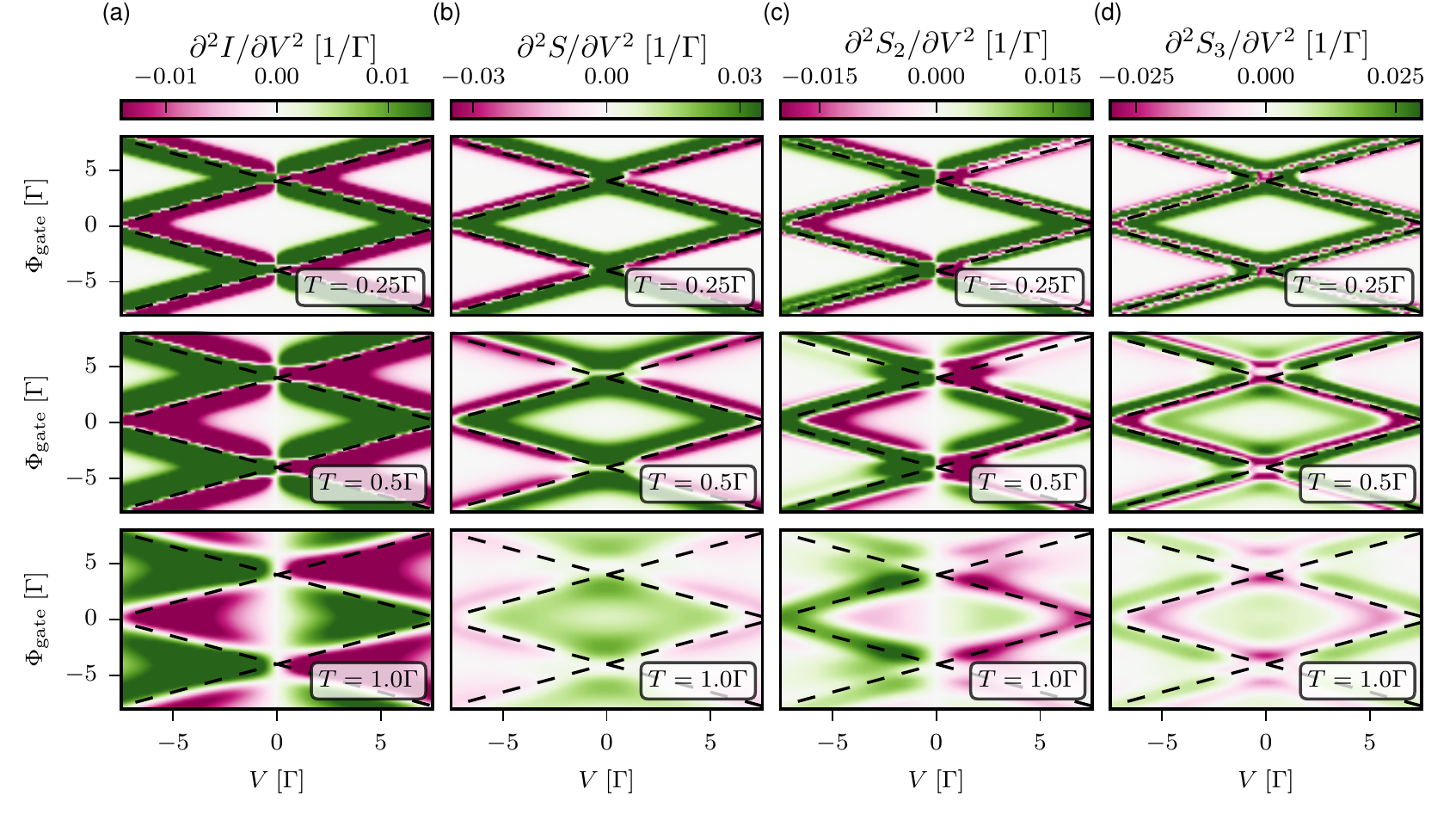}
  \caption{
		QME results.
		The second derivative with respect to bias voltage is shown for the current (a), the noise (b), $S_2$ (c) and $S_3$ (d). From top to bottom, the temperature increases from $T=0.25\Gamma$ to $T=0.5\Gamma$ and finally $T=\Gamma$. 
		The black dashed lines, which serve as a guide for the eye, indicate the conditions $\epsilon_0 = \mu_{\text{L/R}}$ and $2\epsilon_0+U = \mu_{\text{L/R}}$ that separate resonant from nonresonant transport.
		}
  \label{fig:QME_mpas}
\end{figure*}

\begin{figure}[htb!]
  \centering
  \includegraphics{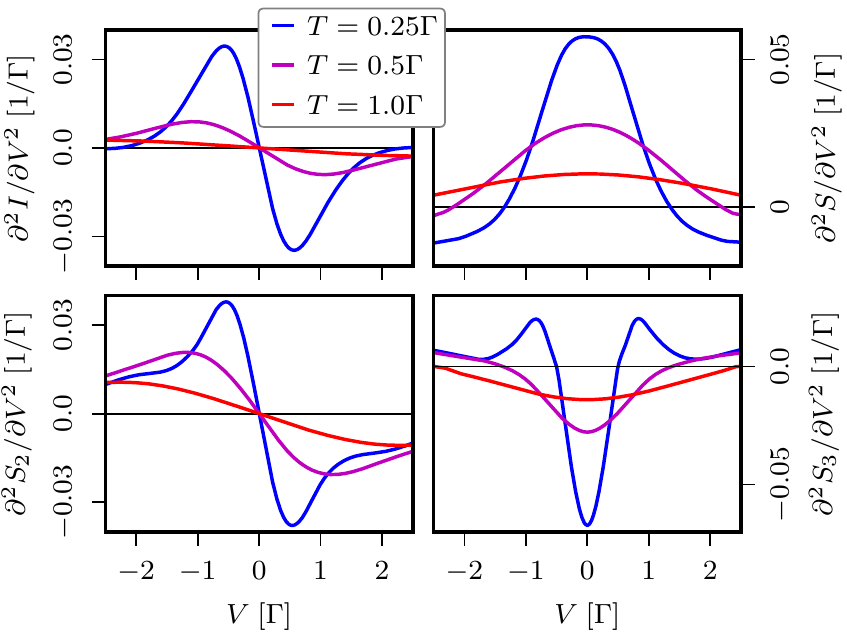}
  \caption{
		NCA results.
		The second derivative with respect to bias voltage is shown for the current $I$ (upper left), the noise $S$ (upper right), and the higher order cumulants $S_2$ (lower left) and $S_3$ (lower right). 
		The gate voltage is set to $\Phi_{\text{gate}}=2\Gamma$, the corresponding Kondo temperature is estimated to be $T_K \approx 0.87\Gamma$.
		These are horizontal cuts across the data in Fig.~\ref{fig:NCA_mpas}, as marked by the red solid lines.
		}
  \label{fig:cuts_ddO_comparison}
\end{figure}

We begin by exploring the influence of Kondo physics on higher order cumulants. 
The aim of this section is to establish the existence of effects in higher order cumulants that are related to the Kondo phenomenon at the edge of the Kondo regime. 
We do not investigate the scaling regime of the Kondo model, where methods like NRG would be most appropriate. 
Rather, we establish that the propagator NCA represents a qualitatively better alternative to QME approximations for the study of counting statistics. 
To this end, we compare NCA results, where a qualitative signature of such phenomena is expected, with QME results, where none is expected.
This is already a huge advantage of the NCA over the QME.
We are working at the edge of the Kondo regime where the dot-lead coupling is not the smallest energy scale of the system. Therefore, agreement between the NCA and the QME data can not be expected, as this would require far higher temperatures, where the temperature is the largest parameter of the system. We refrain from considering such high temperatures in favor of focusing on the installment of the Kondo phenomenon in higher order cumulants.
Later, in Sec.~\ref{sec:NCA_VS_iQMC}, we evaluate that the NCA predictions are more accuracy (yet not qualitative) by comparing with numerically exact iQMC results.

Fig.~\ref{fig:NCA_mpas_dO} provides an overview over the first derivatives with respect to bias voltage of observables $I$, $S$, $S_2$, and $S_3$ as a function of bias and gate voltage calculated by the NCA method.
These first derivatives, such as the conductance $\partial I/\partial V$, are standard observables in various contexts.
Columns of panels correspond to the different observables, while rows correspond to different temperatures.
All derivatives with respect to bias voltage presented in this manuscript are calculated using the symmetric finite difference method on a sufficiently dense grid.
Fig.\ \ref{fig:NCA_mpas_dO} contains a great deal of information in a rather compact form. To make them easier to understand, it is useful to focus on two particular sets of physical features.
First, the transition between resonant and nonresonant transport, which is marked by dashed black lines, and for which agreement between the NCA and the QME results can be found in the large temperature limit.
Second, features associated with the emergence of Kondo and mixed-valence physics are visible in some of the observables. The signature of these correlation-driven effects are features centered around zero bias voltage,
which are more pronounced for some observables than for others, and which disappear with increasing temperature.
Since this central feature and its bias voltage dependency is of primary interest, but can be weak in some regimes, we henceforth consider the second derivative of the cumulants with respect to the bias voltage.

Figs.~\ref{fig:NCA_mpas} and \ref{fig:QME_mpas} provide an overview of the behavior of of the second derivatives $\partial^2 I/\partial V^2$, $\partial^2 S/\partial V^2$, $\partial^2 S_2/\partial V^2$, and $\partial^2 S_3/\partial V^2$ as a function of bias gate voltage at different temperatures in the NCA and QME approximations, respectively.
To facilitate comparison, the figures employ equivalent false color representations of the data.
Yet, we emphasize that in the parameter regime under investigation no agreement between the two approaches can be expected and neither of the two methods is suspected to provide quantitative results.
As before, we predominantly focus on two features, first of which is the transition between the resonant and nonresonant transport regime, which is again highlighted by black dashed lines.
The associated behavior is clearly apparent in all QME plots and accentuated by the fact that the QME method neglects broadening effects provided by the coupling to the leads.
In contrast to that, the NCA method accounts for some broadening provided by the leads, the precise impact of which depends on the parameters of the system as well as the bias and the gate voltage.
This broadening leads to an onset of resonant transport, which is smeared over a wider bias regime as compared to the QME results. This fact is emphasized upon considering the second derivative with respect to bias voltage; even to the extent that some features seen in the QME are completely eliminated by broadening in the NCA.
However, in particular when comparing the NCA and QME data for $T=1.0\Gamma$, where other effects take a backseat, some qualitative agreement between the approaches is observed.

The second feature is the emergence of Kondo physics centered around zero bias voltage, this time clearly visible in the NCA plots, but completely missing from the QME data.
As can be seen by comparing the top and middle panels of Fig.~\ref{fig:NCA_mpas}, higher cumulants reveal progressively richer and more complex dependencies on the bias and gate voltages.
Thus, they provide increasingly detailed modes of characterization.
An interesting point to note is that the temperature at which cumulants exhibit correlated phenomena does not appear to vary significantly with the cumulant order.
This is true both in and out of equilibrium, and to some degree supports the idea that the low energy physics is controlled by a few universal energy scales even when a bias voltage is applied.
Moreover, we notice that the signatures for the Kondo effect appear more pronounced and extend over a larger bias range close to $\Phi_{\text{gate}}=\pm4\Gamma$, in particular when compared to the particle-hole symmetric case at $\Phi_{\text{gate}}=0$.
This can be rationalized with the dependence of the Kondo temperature on the gate voltage, which is estimated to increase from $T_K \approx 0.8\Gamma$ for $\Phi_{\text{gate}}=0$ to $T_K \approx 2.8\Gamma$ for $\Phi_{\text{gate}}=\pm4\Gamma$.\cite{Hewson1997kondo}
Still, we emphasize again that the NCA may described certain trends correctly, but is not expected to give quantitatively reliable results for the Kondo temperature. Also, at the parameter regime under investigation, it can not be expected that the shape of the Kondo features is solely determined by the Kondo temperature.
A more detailed study of these features requires a more systematic study going beyond the NCA, where also the deep Kondo regime and the scaling behavior can be accessed.

Further details are revealed by considering parameters below the resonance condition, at a constant nonzero gate voltage and a range of bias voltages.
A cut of this kind across the data of Fig.~\ref{fig:NCA_mpas} is shown in Fig.~\ref{fig:cuts_ddO_comparison}, and the parameters chosen for the cut are marked in  Fig.~\ref{fig:NCA_mpas} by solid red lines.
We refrain from reproducing the corresponding QME data here, as the QME results does not contain information on the Kondo feature (see Fig.\ \ref{fig:QME_mpas}) and would only complicate the plots.
As even(odd) cumulants are symmetric(antisymmetric) with respect to bias voltage, it is instructive to directly compare $I$ with $S_2$; and respectively $S$ with $S_3$.
The data reveals that $\partial^2I/\partial V^2$ exhibits a single peak--dip structure which corresponds to the well-known peak in conductance at low bias voltage.
Deep in the Kondo regime, the width of the conductance will be given by the Kondo temperature, but we do not expect the NCA to reproduce such physics quantitatively.
Here, at the edge of the Kondo regime, we find that the resonance exhibits additional broadening.
Comparing this to the results for $\partial^2S_2/\partial V^2$, another shoulder appears at low temperature at a bias voltage of about $V\sim1.5\Gamma$.
This indicates that the noise analog of conductance, $\partial S_2/\partial V$, exhibits a structure where two peaks centered around zero bias voltage overlay each other.
The underlying physical mechanism for this feature can not be determined with certainty given the present methodology, but it is known that higher order cumulants are sensitive to a wider energy range and the effect may be associated to the availability of different transport channels at higher bias voltages.
Similarly, $\partial^2S/\partial V^2$ shows a single pronounced peak centered around zero bias voltage, whereas at low temperature, $\partial^2S_3/\partial V^2$ develops distinctive side peaks at a bias voltage $V\sim\Gamma$.
We assume that the width and the magnitude of these features are associated to the Kondo temperature in the deep Kondo regime.
The realization that higher order cumulants show richer Kondo features suggests that they can aid the identification of correlation effects. 
In particular in situations where higher order cumulants are measured but the availability of data is otherwise limited, such that standard procedures like measuring the scaling of the conductance with temperature are not possible.
In many experiments and numerical methods, it is difficult to measure small signals like the current at very low bias voltages. In such cases, considering higher order cumulants may provide a diagnostic tool for identifying Kondo correlations that is applicable at higher biases and is characterized by larger signals.

\subsection{Generalized Fano factors and their implications}\label{sec:Fano}

\begin{figure}[htb!]
\centering
\includegraphics{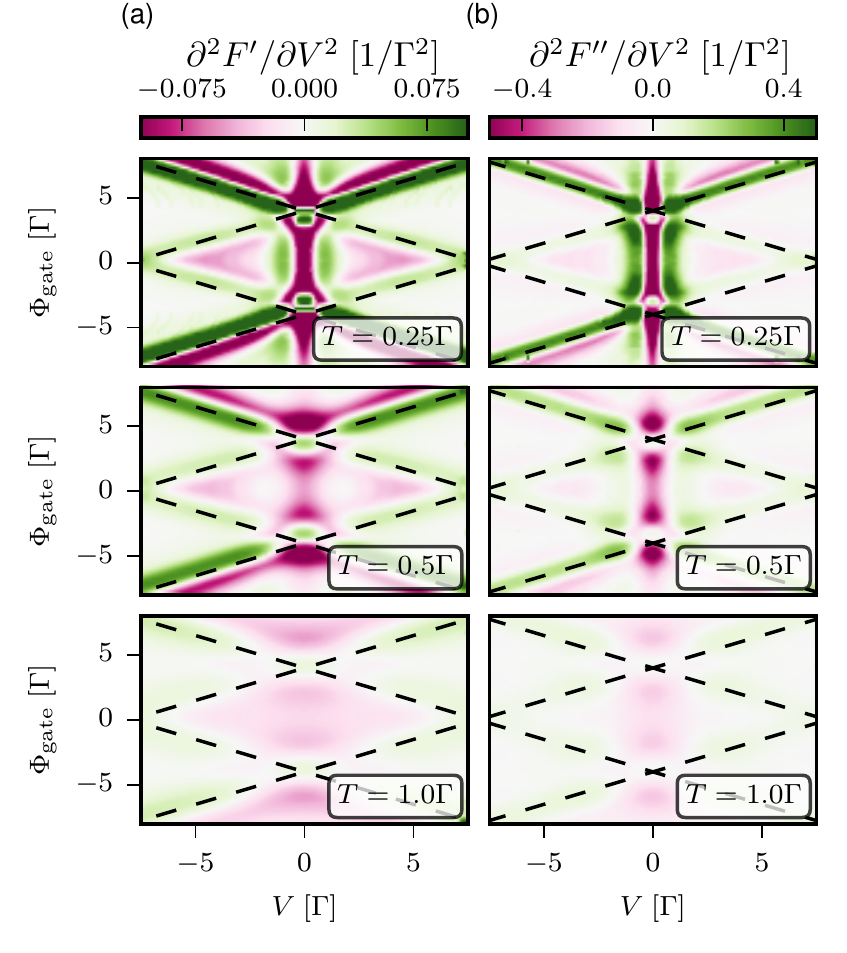}
	\caption{
NCA results.
The second derivative with respect to bias voltage is shown for the generalized Fano factors $F' = S_2/I$ (a) and $F'' = S_3/S$ (b).
From top to bottom, the temperature increases from $T=0.25\Gamma$ to $T=\Gamma$.
The black dashed lines, which serve as a guide for the eye, indicate the conditions $\epsilon_0 = \mu_{\text{L/R}}$ and $2\epsilon_0+U = \mu_{\text{L/R}}$ that separate resonant from nonresonant transport.
}
	\label{fig:ddF_NCA_mpas}
\end{figure}

As noted in Sec.~\ref{sec:FCS}, the Fano factor $F=S/I$ manifests a singularity at zero voltage, where the current (odd with respect to the bias voltage) disappears while the noise (even with respect to the bias voltage) does not.
$F$ will be revisited in Sec.~\ref{sec:NCA_VS_iQMC}, where we benchmark the NCA method against numerically exact results.
In the following, we consider the generalized Fano factors $F' \equiv S_2/I$ and $F'' \equiv S_3/S$.
These are the lowest order ratios comprising only odd and even cumulants, respectively.
They are therefore free of singular behavior at zero voltage, making them potentially useful for exploring Kondo physics.

For both observables, it is once again more convenient to plot the second derivative with respect to bias voltage.
In Fig.~\ref{fig:ddF_NCA_mpas} these are shown at the same parameter ranges used in Figs.~\ref{fig:NCA_mpas} and \ref{fig:QME_mpas}.
$F'$ and $F''$, respectively, are shown in the left and right panels, temperature increases as we go to lower panels.
Both generalized Fano factors exhibit sharp, well defined Kondo features at low temperatures.
As before, these correlation driven features disappear at higher temperatures.

The separate cumulants in Fig.~\ref{fig:NCA_mpas} are dominated by the signature of the transition between off-resonant and resonant transport.
Remarkably, however, in Fig.~\ref{fig:ddF_NCA_mpas} $F'$ exhibits Kondo features of comparable scale to those delineating the resonant transport edge, and $F''$ is dominated by the Kondo features.
This suggests that symmetry-corrected higher order Fano factors contain detailed information regarding correlation effects, and may be a more sensitive probe of such physics than lower order quantities.

As the temperature is lowered and the Kondo effect develops, the value of $F'$ and $F''$ at low bias voltages increases, except near the resonance condition. 
Since the Kondo effect enhances the current $I$, an increase in $F'$ implies that $S_2$ is more strongly enhanced than $I$.
Correspondingly, the underlying probability distribution describing electron transfer becomes increasingly skewed.
Similarly, while the behavior of the noise is more complicated, $S$ is mostly suppressed by Kondo physics, and the same is true for $S_3$.
An increase in $F''$ therefore implies a weaker suppression of $S_3$ than that of $S$, and an increasingly bifurcated probability distribution.
A more detailed analysis of the probabilities $P_L(t,n)$ would be interesting in this regard, but is beyond the scope of the present work.

\subsection{Comparison with numerically exact results}\label{sec:NCA_VS_iQMC}

\begin{figure*}[htb!]
\centering
\includegraphics{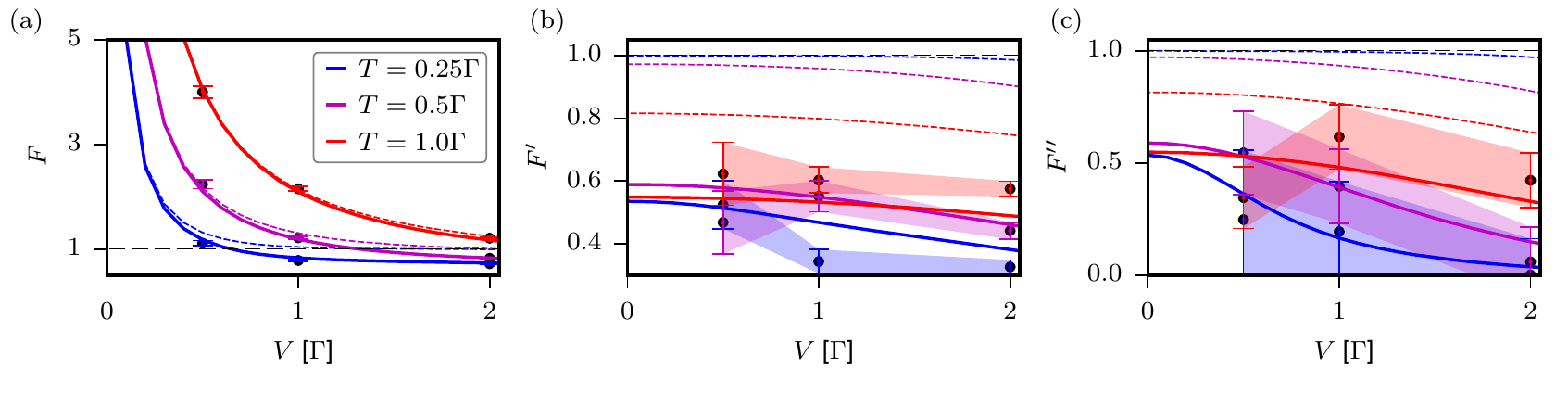}
\caption{
Fano factor $F$ (a); and generalized Fano factors $F'=S_2/I$ (b), and $F''=S_3/S$ (c), at a gate voltage of $\Phi_{\text{gate}}=2\Gamma$.
Colors correspond to different temperatures.
Solid lines are NCA results, dashed lines are QME results, and circles are numerically exact results obtained with iQMC.
The dashed black line in the plots highlights the value $1$, which is associated with a (classical) Poissonian distribution.
}
\label{fig:cuts_F}
\end{figure*}

It is clear from the data that we have presented so far that, when considering higher order transport cumulants, the NCA method captures physics not accounted for by the QME method.
This is not entirely surprising, since it is known to do so for single-particle correlation functions and for the current.
However, since both these techniques are approximate, it is not at all obvious that the NCA actually provides higher accuracy as well.
We will therefore compare the NCA and QME results to numerically exact iQMC data, in order to assess which approximate method provides more accurate results.

Fig.~\ref{fig:cuts_F} depicts the Fano factor $F$ and its generalizations $F'$ and $F''$ as functions of the bias voltage, once again for three different temperatures.
Solid lines represent NCA data and dashed lines represent QME data.
Dots indicate iQMC results converged with respect to all numerical parameters.
Error bars and shading on these dots correspond to confidence intervals (see App.~\ref{app:iQMC_error} for details regarding how these are obtained).
We do not consider second derivatives with respect to the bias voltage here, since obtaining these accurately in iQMC involves further technical challenges.
Similarly, we refrain from discussing data below a bias voltage of 0.5.
We note that in general, lower voltages and higher order cumulants are more difficult to access in iQMC (see Apps.~\ref{app:iQMC_error} and \ref{app:iQMC_lambda}).
Consequently, if one is interested in accessing the details and the scaling of the features discussed above, it might be advantageous to resort to another numerically exact method.

The left panel of Fig.~\ref{fig:cuts_F} shows the Fano factor $F$.
As noted in Sec.~\ref{sec:Fano}, at low bias voltages $F$ is dominated by the Nyquist--Johnson singularity and the isolation of Kondo-related features is difficult, but here we focus on the accuracy of the different methods.
Generally speaking, reasonable agreement can be observed between the NCA, QME and the iQMC results for all temperatures, both qualitatively and quantitatively.
At high temperatures and low voltages, NCA and QME results are almost indistinguishable from each other and accurately capture the trends in the exact result.
Importantly, however, the QME always predicts Poisson statistics with a Fano factor of 1 at large bias voltages.
The NCA correctly captures deviations from this, a result validated by the iQMC data.

Results for the generalized Fano factor $F'$ are presented in the middle panel of Fig.~\ref{fig:cuts_F}.
Overall, the three methods predict a qualitatively similar dependence of $F'$ on bias voltage and temperature, though there are qualitative differences.
The QME method predicts larger values than the NCA approach, while the outcome of the NCA calculations is in better agreement with the iQMC data.
Despite the increased errors associated with the iQMC results for $F'$, it is possible to establish that the NCA method provides more accurate results than the QMC approach.
However, a quantitatively accurate observation, especially at low voltages and temperatures where the Kondo effect can be most cleanly defined and observed, is beyond the iQMC data at hand.

For the second generalized Fano factor $F''$ depicted in the right panel of Fig.~\ref{fig:cuts_F}, the error associated with the iQMC scheme dominates the exact data to the extent that trends in the bias and temperature dependence are non obvious. For this Fano factor, the iQMC method in its current implementation breaks down, indicating an area where the usage of approximate schemes is more favorable.
When comparing the QME and the NCA results, the QME approach again predicts larger values for $F''$ than the NCA method. As before, the NCA data is in better agreement with the iQMC results, hinting towards a higher accuracy of the NCA method. For a more detailed analysis, better iQMC data is required.

\section{Summary}\label{sec:summary}
We developed a simple theoretical approach based on the noncrossing approximation (NCA) to the study of full counting statistics (FCS) in nonequilibrium transport, and implemented it for the Anderson impurity model.
The approach can be easily generalized to more generic models.
Its accuracy can be improved by diagrammatic means, for example by considering one-crossing and vertex corrections.
The NCA method requires substantially more modest computational resources than its numerically exact counterpart, the inchworm Monte Carlo (iQMC) method; and is for most practical purposes almost as easy to use as the commonly employed quantum master equations (QMEs).
In the present case, the QME and NCA data was generated on a desktop workstation within a few hours and approximately a day, respectively; while the iQMC results were generated over several days on a small cluster.
Despite this simplicity, the NCA captures some physics not present in the QME approximation.

To showcase the advantages of the NCA approach to FCS, we compared it against the QME method for the first few transport cumulants.
Unsurprisingly, this illustrated that the first shows signatures of the Kondo effect while the latter does not.
More interestingly, it showed that the NCA predicts a rich and detailed set of features in the higher order cumulants.

Experimentally, it is often advantageous to consider ratios between transport cumulants, like the Fano factor.
However, at low bias voltages the Fano factor is dominated by a Nyquist--Johnson singularity that obstructs one's view of Kondo-related features.
We explored a set of generalized, symmetry-motivated Fano factors constructed from higher order cumulants that are designed to remove this singularity.
Within the NCA method, we showed that these quantities embody excellent probes of Kondo physics.

Finally, we established the accuracy of the method upon comparison with  numerically exact benchmarks obtained from the iQMC scheme.
We showed that the predictability of approximate NCA method is superior to data provided by the QME approach. For the Fano factor, we demonstrated that the NCA can even provide qualitative results.

\section*{Acknowledgements}
A.E. was supported by the Raymond and Beverly Sackler Center for Computational Molecular and Materials Science, Tel Aviv University.
E.G. was supported by the Simons Collaboration on the Many Electron Problem.
G.C. acknowledges support by the Israel Science Foundation
(Grants No.~1604/16 and 218/19).
This research was supported by Grant No.~2016087 from the United States-Israel Binational Science Foundation (BSF).

\appendix
\section{Calculating observables associated with higher order cumulants within the iQMC scheme}\label{app:iQMC_error}

\begin{figure}[htb!]
	\includegraphics{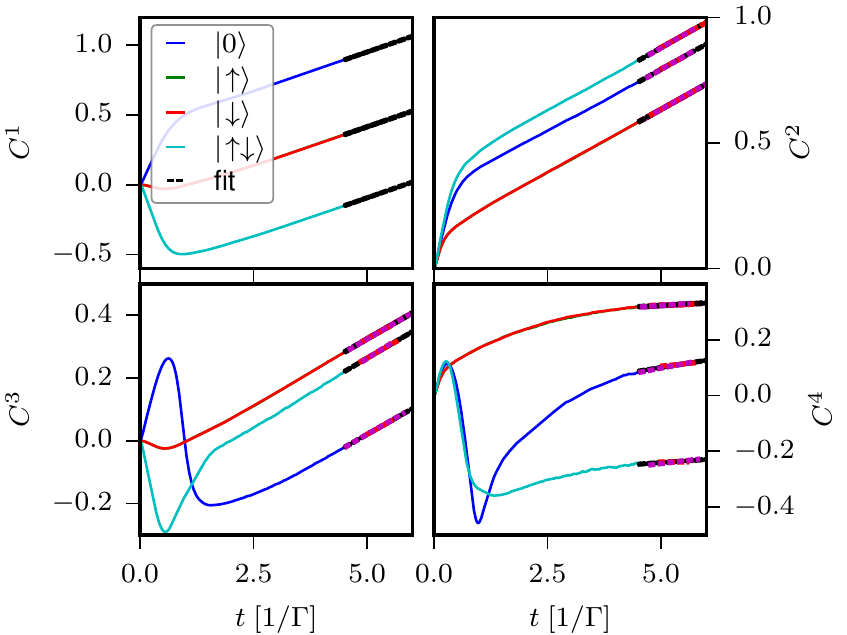}
	\caption{
		Visualization of the procedure for determining the observables and the correspond errors from iQMC data.
		The representation data set shown here corresponds to $T=0.25$ and $V=1$.
		The finite counting field used for the calculation is $\lambda=0.75$.
		The individual panels show the behavior of the different cumulants as a function of time. The different lines correspond to different initial conditions, the fits are marked by dashed lines.
	}
	\label{fig:iQMC_errors}
\end{figure}

In this appendix, we describe how the observables $I = \frac{\partial}{\partial t} C^1$, $S=\frac{\partial}{\partial t} C^2$, $S_2 = \frac{\partial}{\partial t} C^3$, and $S_3 = \frac{\partial}{\partial t} C^4$ were calculated for the steady state within the iQMC framework.
We also provide an account of how confidence intervals are estimated.
Whereas the main goal of this paper is to introduce the NCA methodology and its advantages to theorists and experimentalists interested in transport counting statistics, this appendix is aimed at more specialized readers interested in numerically exact methodologies like the iQMC.

In steady state, all cumulants increase linearly with time.
We simulate time propagation using the iQMC method until the cumulants display a linear dependency on time.
Then, we perform a linear fit to this part of the data. Using the outcome of the fitting routine and averaging over all possible initial conditions, we determine the observables. This procedure is visualized in Fig.\ \ref{fig:iQMC_errors} for a representative data set. Fig.\ \ref{fig:iQMC_errors} also allows for an assessment of the intrinsic noise of the iQMC method and the requirement of a linear behavior of the respective cumulants.
The error for the iQMC data is then estimated upon considering extremal values for lines connecting two  points within the time range where the cumulants display a linear dependence with time (different colored dashed lines in Fig.\ \ref{fig:iQMC_errors}). Again, we average over all initial conditions, whereby we perform a Gaussian error propagation. Strictly speaking, this approach suffers from the fact that the different initial conditions are not statistically independent. Still, the scheme provides a reasonable estimate for the underlying error. 

In contrast to the iQMC method, the approximate NCA and QME approaches employed in this work do not exhibit a statistical error. As such, the derivatives entering the expressions for the observables can be calculated by means of finite differences and no fitting procedure is required. Again, the system is propagated until the steady state establishes.

\section{Counting field dependence of iQMC data}\label{app:iQMC_lambda}
\begin{figure}[tb!]
	\includegraphics{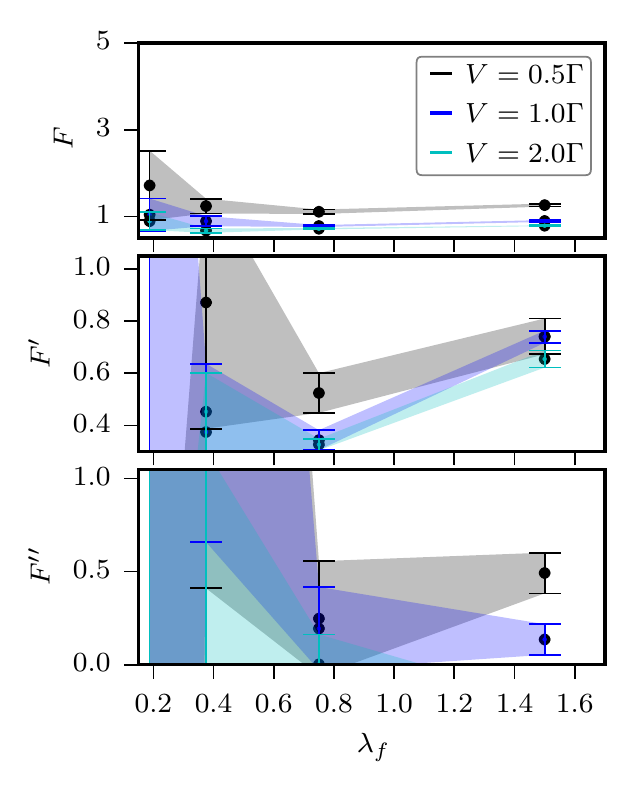}
	\caption{
		Visualization of the dependence of the iQMC data on the finite counting field $\lambda_f$ used for the numerical calculation for different bias voltages $V$.
		The representation data set shown here corresponds employs to $T=0.25\Gamma$.
		The individual panels show the behavior of the different Fano factors discussed in Sec.\ \ref{sec:NCA_VS_iQMC}. 
		The different lines correspond to different bias voltages, the shaded areas indicate an estimate for the underlying error.
		For reference, the scaling of the individual panels corresponds to the scaling of the associated plots in Fig.\ \ref{fig:cuts_F}.
	}
	\label{fig:iQMC_lambda}
\end{figure}
The cumulants within the FCS framework are given by the derivatives of the logarithm of the generating function $Z(t, \lambda)$ with respect to the counting field $\lambda$ at $\lambda=0$ (see Eq.~(\ref{eq:def_cumulant})). For numerical applications, the generating function is determined for finite values of the counting field and the derivative with respect to $\lambda$ is calculated, for example, by means of finite differences.

For the iQMC scheme, the statistical error associated with the different cumulants depends on the finite value employed for the counting field. 
Within this work, we calculate the derivative by means of symmetric finite differences, 
\begin{subequations}
\begin{eqnarray}
C^1(t)	&=&	-i\frac{\partial}{\partial\lambda} \ln\left(Z(t, \lambda)\right) \Big|_{\lambda=0}  \\
        &\approx& \frac{\ln(Z(t, \lambda_f)) - \ln(Z(t, -\lambda_f))}{2\lambda_f} ,\nonumber\\
C^2(t)	&=&	-\frac{\partial^2}{\partial\lambda^2} \ln\left(Z(t, \lambda)\right) \Big|_{\lambda=0} \\
        &\approx& \frac{\ln(Z(t, \lambda_f)) - \ln(Z(t, 0)) + \ln(Z(t, -\lambda_f))}{\lambda_f^2}, \nonumber
\end{eqnarray}
\end{subequations}
etc.\  for the finite counting field $\lambda_f$.
The dependence of the (generalized) Fano factors studied in Sec.~\ref{sec:NCA_VS_iQMC} on the finite counting field $\lambda_f$ used for calculating the derivative is visualized in Fig.~\ref{fig:iQMC_lambda}. For large values of $\lambda_f$, the error is small but the estimate for the derivative provided by the numerical derivative deviates from the true value. With decreasing $\lambda_f$, the error associated with the iQMC data increases. Moreover, as higher order cumulants depend on higher order derivatives with respect to the counting field, the numerical error increases and determining accurate data for quantities depending on higher order cumulants becomes increasingly challenging.

We mention that there are also other possible approaches to determine the derivative with respect to the counting field. As such, an alternative but more expensive route is to obtain the full FCS and calculate the generating function $Z(t, \lambda)$ for various different values of $\lambda$. In this case, the derivatives can be determined analytically for a polynomial fit for the counting field dependence of the generating function. This approach was employed, for example, in Ref.\ \onlinecite{Ridley2018}.

\bibliography{bib}

\end{document}